\documentclass[12pt,preprint]{aastex}
\usepackage{graphicx}

\title{A Blazar-Like Radio Flare in Mrk\,231}
\author{Cormac Reynolds\altaffilmark{1}, Brian Punsly\altaffilmark{2}, Christopher P. O'Dea\altaffilmark{3}, Natasha Hurley-Walker\altaffilmark{1}}
\altaffiltext{1}{ICRAR-Curtin University, GPO Box U1987, Perth,
Western Australia, 6102, Australia} \altaffiltext{2}{1415 Granvia
Altamira, Palos Verdes Estates CA, USA 90274 and ICRANet, Piazza
della Repubblica 10 Pescara 65100, Italy, brian.punsly1@verizon.net
or brian.punsly@comdev-usa.com} \altaffiltext{3}{Laboratory for
Multiwavelength Astrophysics, School of Physics and Astronomy,
Rochester Institute of  Technology, 54 Lomb Memorial
Drive,Rochester, NY 14623}

\begin{document}

\begin{abstract}
Radio monitoring of the broad absorption line quasar (BALQSO)
Mrk\,231 from 13.9~GHz to 17.6~GHz detected a strong flat spectrum
flare. Even though BALQSOs are typically weak radio sources, the
17.6~GHz flux density doubled in $\approx 150$~days, from $\approx
135$~mJy to $\approx 270$~mJy. It is demonstrated that the elapsed
rise time in the quasar rest frame and the relative magnitude of the
flare is typical of some of the stronger flares in blazars that are
usually associated with the ejection of discrete components on
parsec scales. The decay of a similar flare was found in a previous
monitoring campaign at 22~GHz. We conclude that these flares are not
rare. The implication is that Mrk\,231 seems to be a quasar in which
the physical mechanism that produces the BAL wind is in tension with
the emergence of a fledgling blazar.
\end{abstract}

\section{Introduction}

One of the main mysteries of the quasar phenomenon is the associated
powerful outflows that come in a variety of forms. These outflows
can be manifest as extremely energetic relativistic jets $> 100$~kpc
in extent or massive broad absorption line (BAL) winds. Furthermore,
the existence of large scale jets and BAL winds are almost mutually
exclusive. The propensity for suppressed large scale emission
increases strongly with BALnicity index
\citep{bec00,bec01}\footnote{We use the original definition of a BAL
as UV absorbing gas that is blue shifted at least 5,000 km/s
relative to the QSO rest frame and displaying a spread in velocity
of at least 2,000~km~s$^{-1}$, \citep{wey91}. Note that this
definition specifically excludes the so-called ``mini-BALQSOs," with
the BALNicity index = 0 \citep{wey97}. This is desirable since the
DR5 statistical analysis of \citet{zha10} indicate that these types
of sources (mini-BALQSOs have a large overlap in definition with the
intermediate width absorption line sources of \citet{zha10}) tend to
resemble non-BALQSOs more than BALQSOs in many spectral properties.
Mini-BALQSOs also tend to have much smaller X-ray absorbing columns
than BALQSOs \citep{pun06}. Typically, ``BALQSO" radio targets are
actually mini-BALQSOs since they have larger radio fluxes than
bona-fide BALQSOs, e.g. \citet{bru13, hay13}.}. From Very Long
Baseline Array (VLBA) studies of the BAL quasar (BALQSO) Mrk\,231 in
\citet{rey09} and other radio quiet quasar studies, it has become
evident that radio quiet active galactic nuclei (AGN) can have
relativistic outflows with significant kinetic luminosity, but
possibly only for short periods of time \citep{bru00,blu03}. So this
raises the question what is it that makes some sources radio quiet
and others radio loud? Does the BAL wind inhibit the efficacy of jet
initiation or does it simply limit the ability of a jet to propagate
to large distances, or both? At a redshift of $z = 0.042$, Mrk\,231
is one of the nearest radio quiet quasars to earth. The radio core
is perhaps the brightest of any radio quiet quasar (and certainly
the brightest BALQSO core) at high frequency (22 and 43~GHz).
Studying the radio core at high frequency can provide vital clues to
the origin of both the large scale radio jets and the BAL winds in
AGN. Thusly motivated, the authors have embarked on a program of
high frequency VLBA observations \citep{rey09} and long term high
frequency, densely time sampled, low resolution radio monitoring. We
report our first four years of radio monitoring results here.

\par The paper is organized as follows. Section~2 will describe
previous evidence of the blazar-like nature of Mrk\,231. This is the
motivation for expected dramatic behavior in the high frequency
light curve. The next section describes the observational details of
our monitoring. In Section~4, we compare our most recent epoch of
monitoring to typical strong blazar flares.
Throughout this paper, we adopt the following cosmological
parameters: $H_{0}$=71 km s$^{-1}$ Mpc$^{-1}$,
$\Omega_{\Lambda}=0.73$ and $\Omega_{m}=0.27$.

\section{Previous Observed Blazar-Like Behavior}

The most striking finding in our summary of the VLBA observation in
\citet{rey09} was the strong 22~GHz flare that emerged from the core
between epochs 2006.07 and 2006.32 ($>150\%$ increase in less than 3
months). All attempts in \citet{rey09} to model the high frequency
peak of the spectral turnover, 19.5~GHz, in combination with the
steep spectral index\footnote{for spectral index we use $S_\nu
\propto \nu^{-\alpha}$ throughout.} above 22~GHz ($\alpha \approx
2$) indicate that the flare is synchrotron self-absorbed and the
brightness temperature is $T_{B} \approx 10^{12}\,\mathrm{K}$,
unless the flux density is Doppler boosted. The Doppler boosted
models indicated an intrinsic (rest frame of the plasma) brightness
temperature $\gtrsim 10^{11}\,\mathrm{K}$. The modeling of the flare
in \citet{rey09} also requires that the bulk of the flare emission
is from a region on the order of  $3 \times 10^{16}$~cm.

\par
\begin{figure}
\begin{center}
\includegraphics[width= 0.48\textwidth]{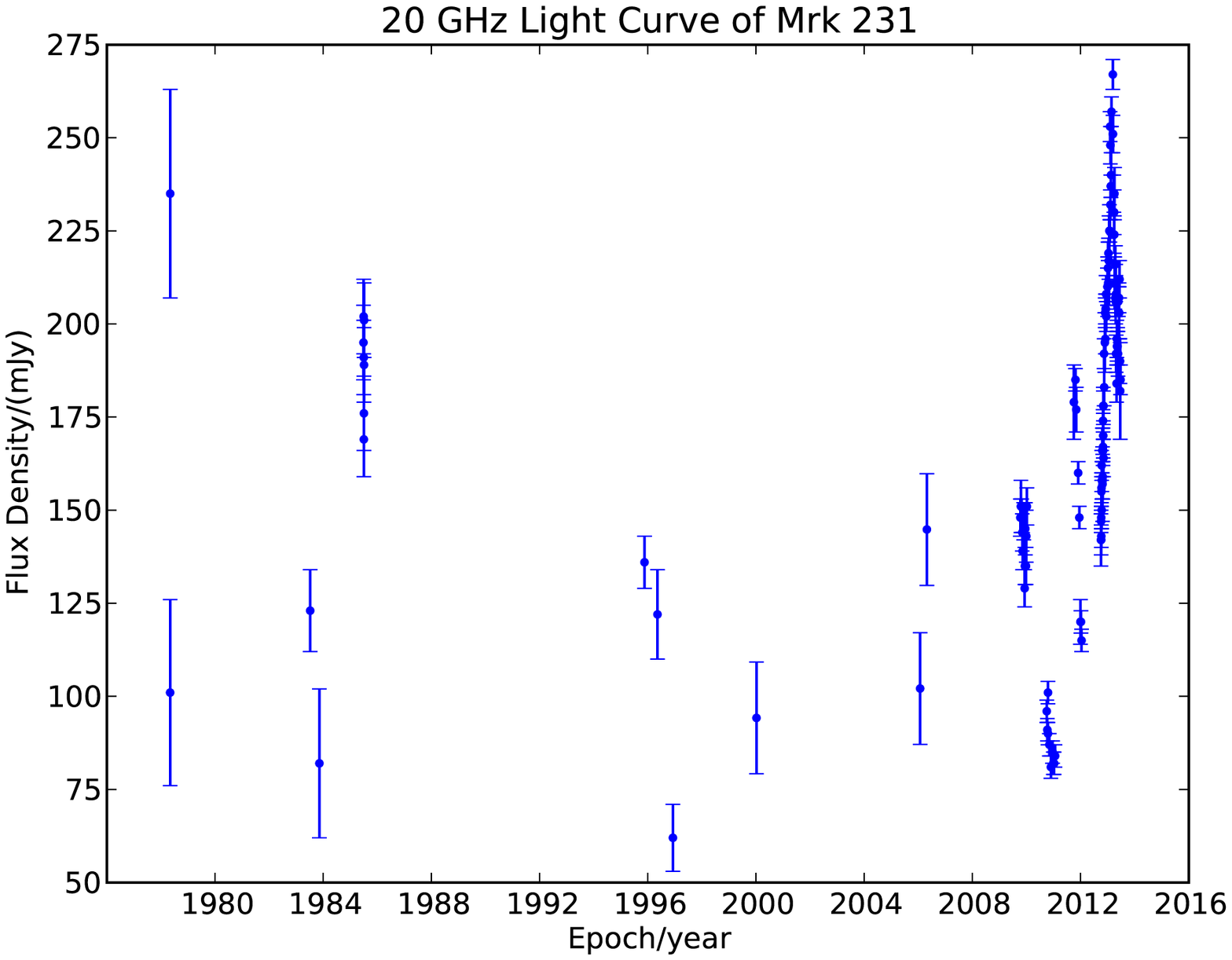}
\includegraphics[width= 0.48\textwidth]{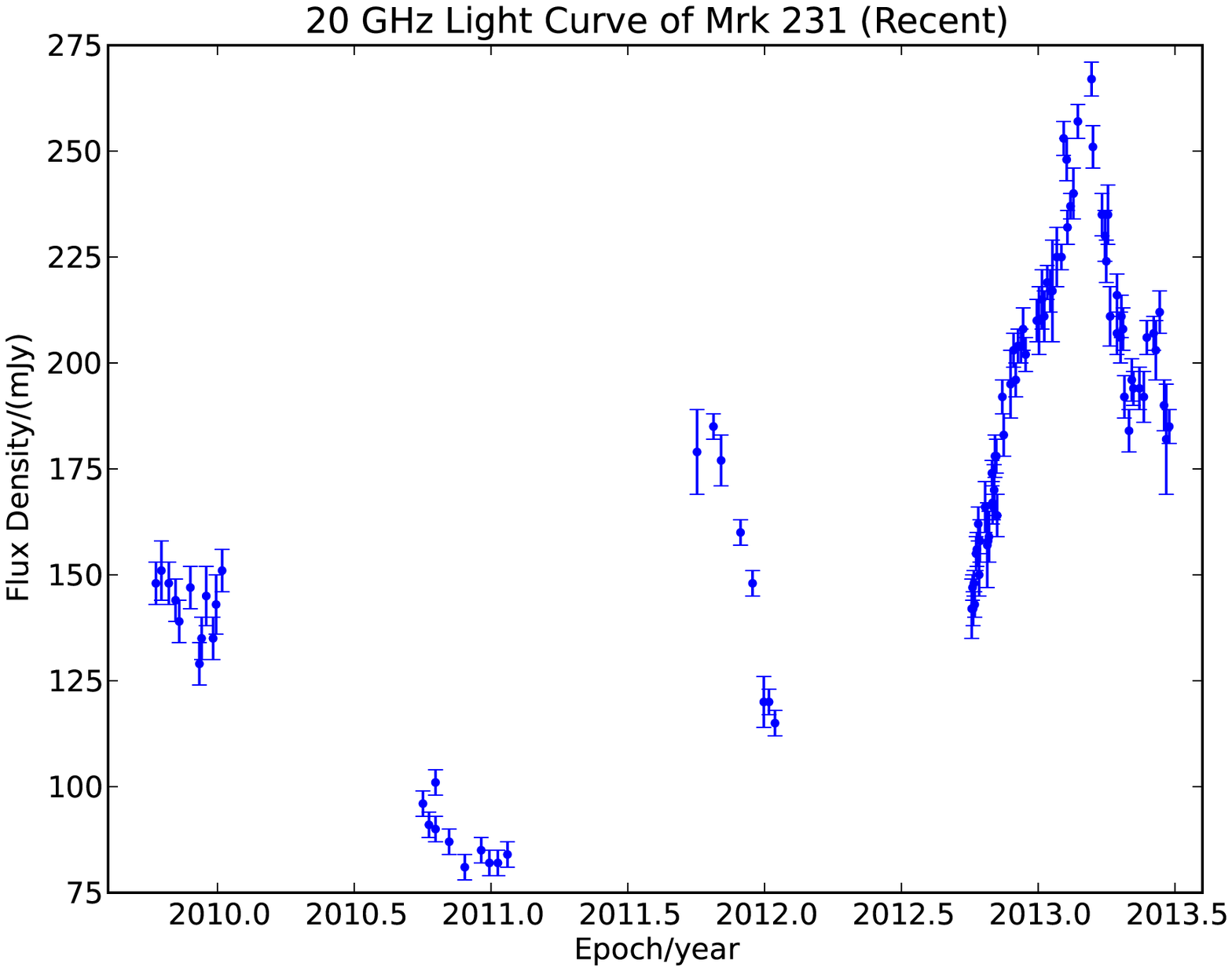}

\caption{The left panel shows the long term, 35 year, light curve at
$\sim 20$~GHz. The right hand panel is a zoom in on the more recent
data, with its much more frequent sampling. \label{fig:lightcurve} }
\end{center}
\end{figure}

 There is strong corroboration in the literature of this blazar-like
behavior. A $134 \pm 38$~mJy flux density variation in 1~day at
22.2~GHz was reported in \citet{mcc78}. This can be seen in
Figure~\ref{fig:lightcurve}. Using the methods of \citet{gho07}, the
time variability brightness temperature was found in \citet{rey09}
to be $T_{B} = (12.4 \pm 3.5)\times 10^{12}$~K. If the brightness
temperature ($T_{B}$) exceeds $10^{12}$~K, it requires a nearly
pole-on orientation and relativistic motion for the jet in order to
avoid the ``inverse Compton catastrophe" \citep{mar79}. This
indicates that the line of sight to the jet is restricted
kinematically to $\theta_{max} <
(25.6^{\circ})^{+3.2^{\circ}}_{-2.2^{\circ}}$ \citep{rey09}.

\section{The Radio Observations}

\subsection{Observations and Calibration}

The program utilized long term monitoring with the VLA (Very Large
Array) and EVLA (Expanded Very Large Array)  at 22~GHz and AMI
\citep[Arcminute Microkelvin Imager;][]{zwa08} at 13.5 -- 18~GHz
\footnote{The Arcminute Microkelvin Imager consists of two radio
interferometric arrays located in the Mullard Radio Astronomical
Observatory, Cambridge, UK \citep{zwa08}. Both arrays observe
between 13.9 and 18.2\,GHz in six frequency channels. The Small
Array (AMI-SA) consists of ten 3.6 m diameter dishes with a maximum
baseline of 20\,m, yielding an angular resolution of 3\arcmin, while
the Large Array (AMI-LA) comprises eight 12.6 m diameter dishes with
a maximum baseline of 110\,m, giving an angular resolution of
0\farcm5.}. These data are plotted in Figure~\ref{fig:lightcurve}
along with historical data both from the literature
\citep{mcc78,ede87,ulv99,ulv00,rey09} and from the VLA public
archive (project codes AB783, AN030, AU015). Both our monitoring
data and the data from the VLA archive were calibrated using NRAO's
AIPS package via the ParselTongue interface \citep{ket06} in the
standard way. 3C\,286 served as the primary flux calibrator in all
the VLA observations. In our VLA/EVLA monitoring observations (2009
-- 2012) we also made use of J1219+4829 as a nearby secondary flux
calibrator. Flux density error estimates on Mrk\,231 were confirmed
by short scans on another VLA calibrator (J1400+6210) located at a
similar separation from the secondary calibrator as Mrk\,231 and
calibrated in an identical fashion to Mrk\,231. Both AMI's Large
Array and Small Array were used, but the Small Array data are
preferred, as explained below. The AMI data were calibrated using
the standard AMI pipeline.

\subsection{Constructing the Historical 20~GHz Lightcurve \label{sect:lightcurve}}

Mrk\,231 appears to have no significant 20~GHz emission on scales
larger than an arcsecond (it appears as a point source to the VLA in
all configurations), so the wide range in spatial resolution
provided by the various instruments in the historical observations
of this source has little effect on the measured flux densities. The
exceptions are the VLBI measurements which resolve out some of the
larger scale, presumably diffuse, emission. In order to convert the
VLBA data in \citet{rey09} to that which would have been detected
with an array containing shorter baselines (such as the VLA), we
separate out the steady background components. In 1996.93 the VLA
measured $62 \pm 9 $~mJy at 22~GHz and the nuclear double was
observed simultaneously with the VLBA to be $\sim 30$~mJy
\citep{ulv99}. We designate this as
$S_{\nu=22\mathrm{GHz}}(\mathrm{wide\; field}) \approx 30$ mJy,
which must be added to the VLBA measurement flux density to get the
total flux density at 22~GHz (i.e., the 2006 data points in
Figure~\ref{fig:lightcurve}). The scientific interest here is to
detect the flaring core on the background of the quasi steady
component. Three separate VLBA observations resolve the nuclear
double at 22 GHz \citep{rey09}. The secondary flux density,
$S_{\nu=22\mathrm{GHz}}(\mathrm{secondary})$, is fairly steady
ranging from 36 to 43~mJy. Thus, to find the core flux density from
the total flux density at any epoch
\begin{eqnarray}
&& S_{\nu=22\mathrm{GHz}}(\mathrm{core}) =
S_{\nu=22\mathrm{GHz}}(\mathrm{total}) -
S_{\nu=22\mathrm{GHz}}(\mathrm{secondary})-
S_{\nu=22\mathrm{GHz}}(\mathrm{wide\; field})\\ \nonumber
&&=S_{\nu=22\mathrm{GHz}}(\mathrm{total}) - 70 \mathrm{mJy} \pm 10
\mathrm{mJy}\;.
\end{eqnarray}

\par The first two monitoring efforts were with the VLA at 22~GHz in
the final quarters of 2009 and 2010 (project codes AR699 and AR717).
In 2009, the flux density was steady and slightly elevated relative
to the long term average. In 2010, the flux density was steady again
and slightly suppressed relative to the long term average. The third
monitoring campaign utilized the EVLA at 22~GHz (VLA11B-019) and was
executed between October 2011 and January 2012. We detected the
decay of what must have been a very strong flare earlier in 2011. In
the last quarter of 2012, we switched to AMI for monitoring which is
continuing. The advantage of AMI is that it can provide more
frequent monitoring with a lower calibration overhead than the EVLA.
The disadvantage is that the maximum available frequency is only
18~GHz, (total useful frequency range 13.5 -- 18~GHz).

\par
\begin{figure}
\begin{center}
\includegraphics[width= 0.7\textwidth ]{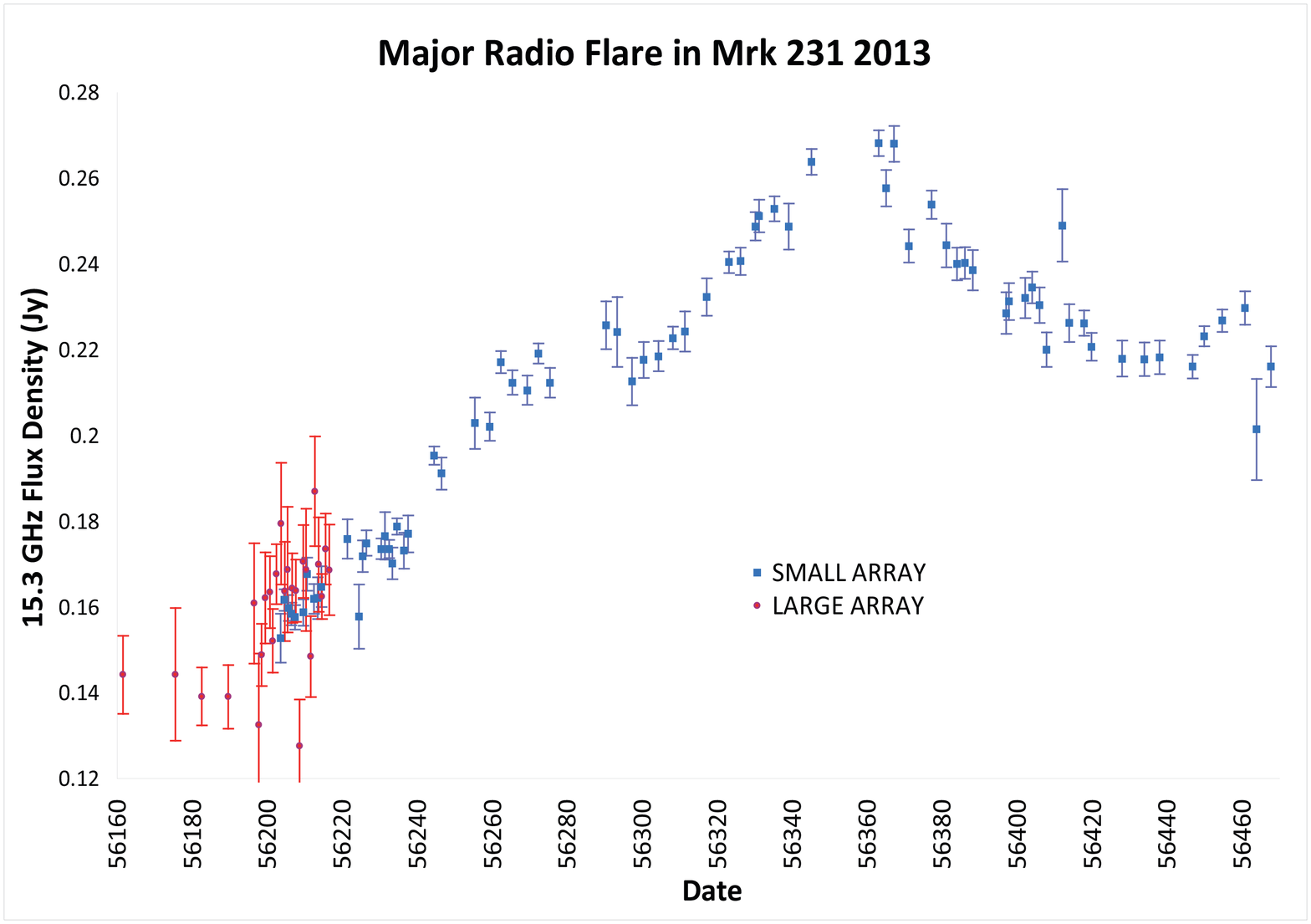}
\includegraphics[width= 0.7\textwidth ]{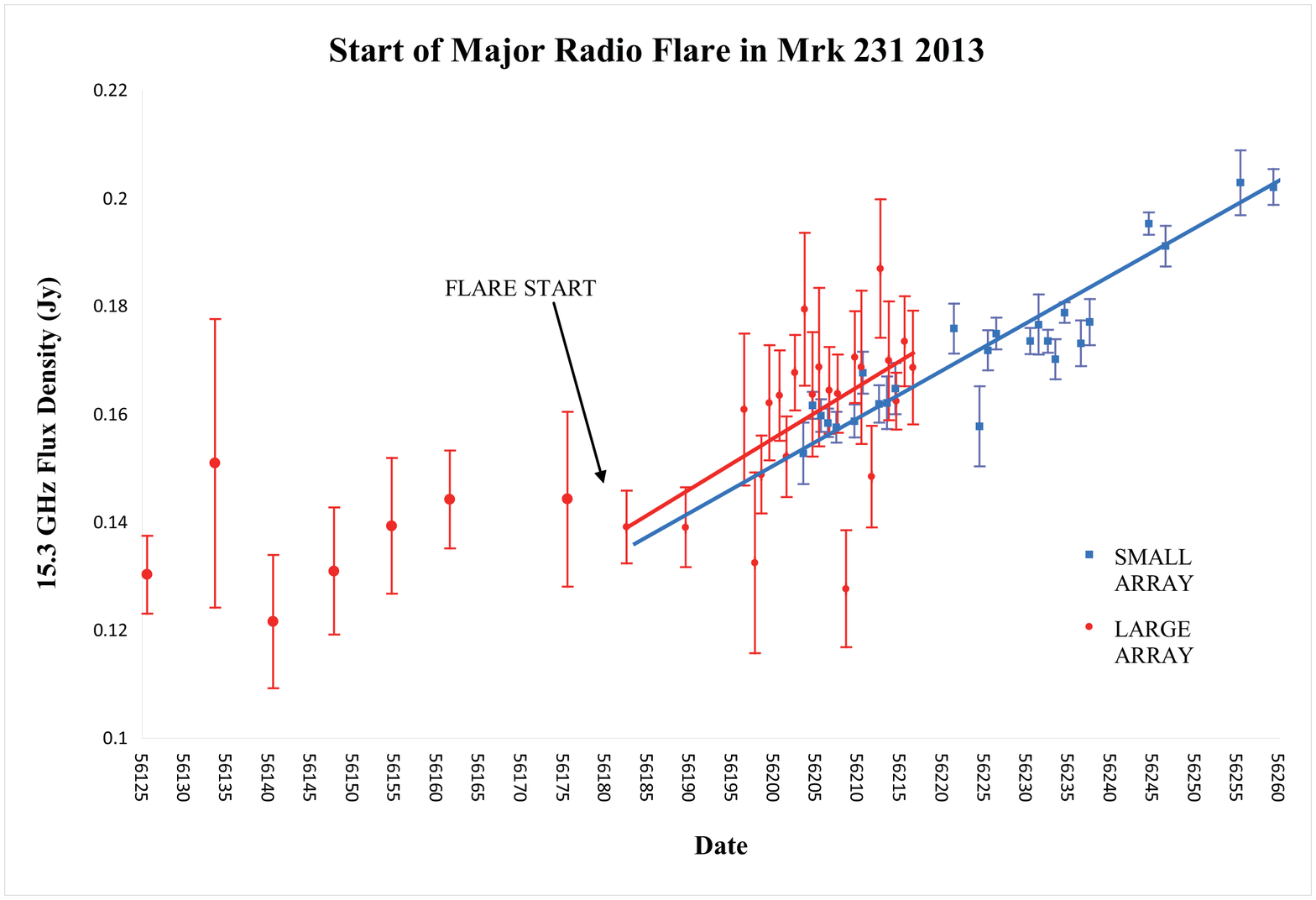}
\caption{The top panel of the figure shows the AMI light curve at
15.3 GHz (a traditional AGN monitoring frequency). The similar flux
density levels to those of the 17.6 GHz light curve in Figure 1 is
indicative of the fact that the spectrum was flat at most epochs.
The light curve includes both AMI-LA data and SA data. The close-up
view in the bottom frame is used to estimate the beginning of the
flare. Notice that in spite of the systematic errors in the LA data,
the linear fit to the data (the red line) yields almost the same
start time as the linear fit to the SA data (the blue line). The
$\sim 0- 2$ day difference is insignificant compared to the $\sim$
150 day rise time of the flare.\label{fig:ami} }
\end{center}
\end{figure}

 Figure~\ref{fig:ami} shows the 15.3~GHz light curve from the 8~months of
monitoring with AMI. The observations began with the AMI Large Array. However,
we noticed flux density variations that exceeded the formal error estimates
from the AMI pipeline but seemed unlikely to be due to source variability. In
order to test this hypothesis we began taking simultaneous measurements with
the Small Array. The Small Array data appeared much more consistent. We
conjecture that the Large Array variations were associated with a high
sensitivity to weather conditions which could not be properly calibrated. After
realizing this, we switched to only Small Array monitoring. Considering the
magnitude of the flare, the accuracy of the Large Array data is adequate as a
whole for finding the start of the flare, although any measurement taken
individually would be suspect. This is verified by taking the linear fit to the
Small Array data near the overlap region and extending it to earlier times in
the bottom frame of Figure~\ref{fig:ami}. The flare start is approximately MJD
56185 -- 56190.

\par We only show the AMI data at one frequency in
Figure~\ref{fig:ami} for the sake of clarity since the spectrum is
flat and the data points overlap. Typically the spectral index is
about $\alpha =0.1 - 0.2$. However, the data are formally consistent
with $\alpha = 0$. The flare in Figure~\ref{fig:ami} seems to have a
magnitude and decay time similar to that indicated by the declining
flux density seen in our EVLA observations in the last quarter of
2011 (see Figure~\ref{fig:lightcurve}). The magnitude of this flare
is larger than the highest previously measured flux density, $235
\pm 28$~mJy in 1976 at 22.2 GHz \citep{mcc78}. On MJD~56363, the
17.6~GHz flux density was $267 \pm 4$~mJy with a spectral index
$\alpha \approx 0.07$, the corresponding spectral energy, was $\nu
L_{\nu} =1.9 \times 10^{41} \mathrm{ergs/s}$.

\section{Comparison to Blazar Flares}
Given the blazar-like properties of Mrk\,231 noted in Section~2, we
compare the strong flare in the 17.6~GHz light curve from AMI to
archival 22~GHz blazar flare light curves in this section. Large
blazar flares are typically associated with the ejection of
components from the nucleus that can be detected on parsec scales
with VLBI. Famous examples include 3C\,273 and BL~Lacertae
\citep{abr96,kir90,mut90,tat99}. However, counter-examples exist
when the core brightens, yet no ejected component is resolved with
VLBI \citep{sav02}. We pick the frequency of 22~GHz since it is high
enough that the total flux density will be dominated by the flat
spectrum radio core. Furthermore, a large data set at 22~GHz of
blazar light curves can be found in \citet{ter04}.

\begin{figure}
\begin{center}
\includegraphics[width= 0.8\textwidth ]{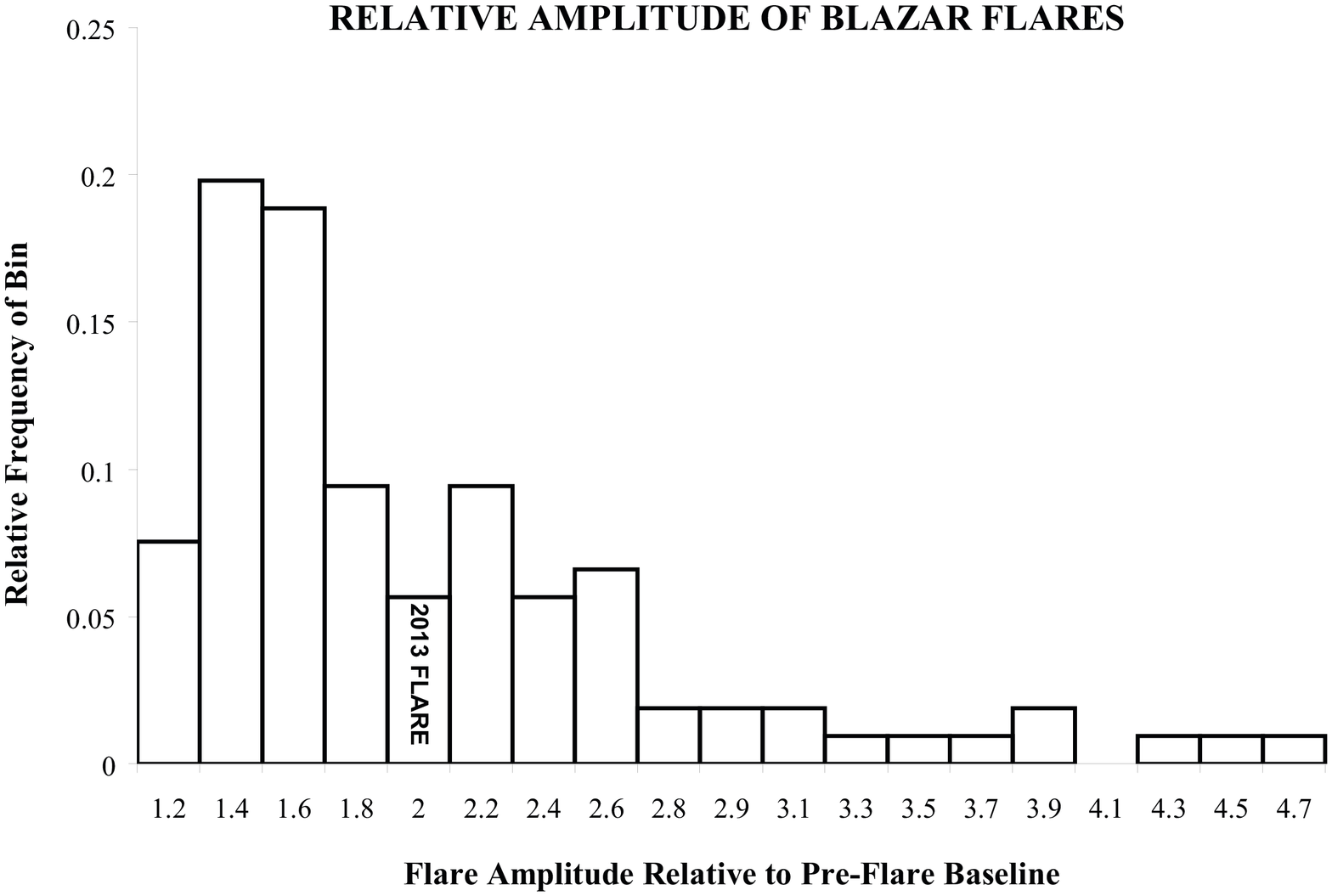}
\includegraphics[width= 0.8\textwidth ]{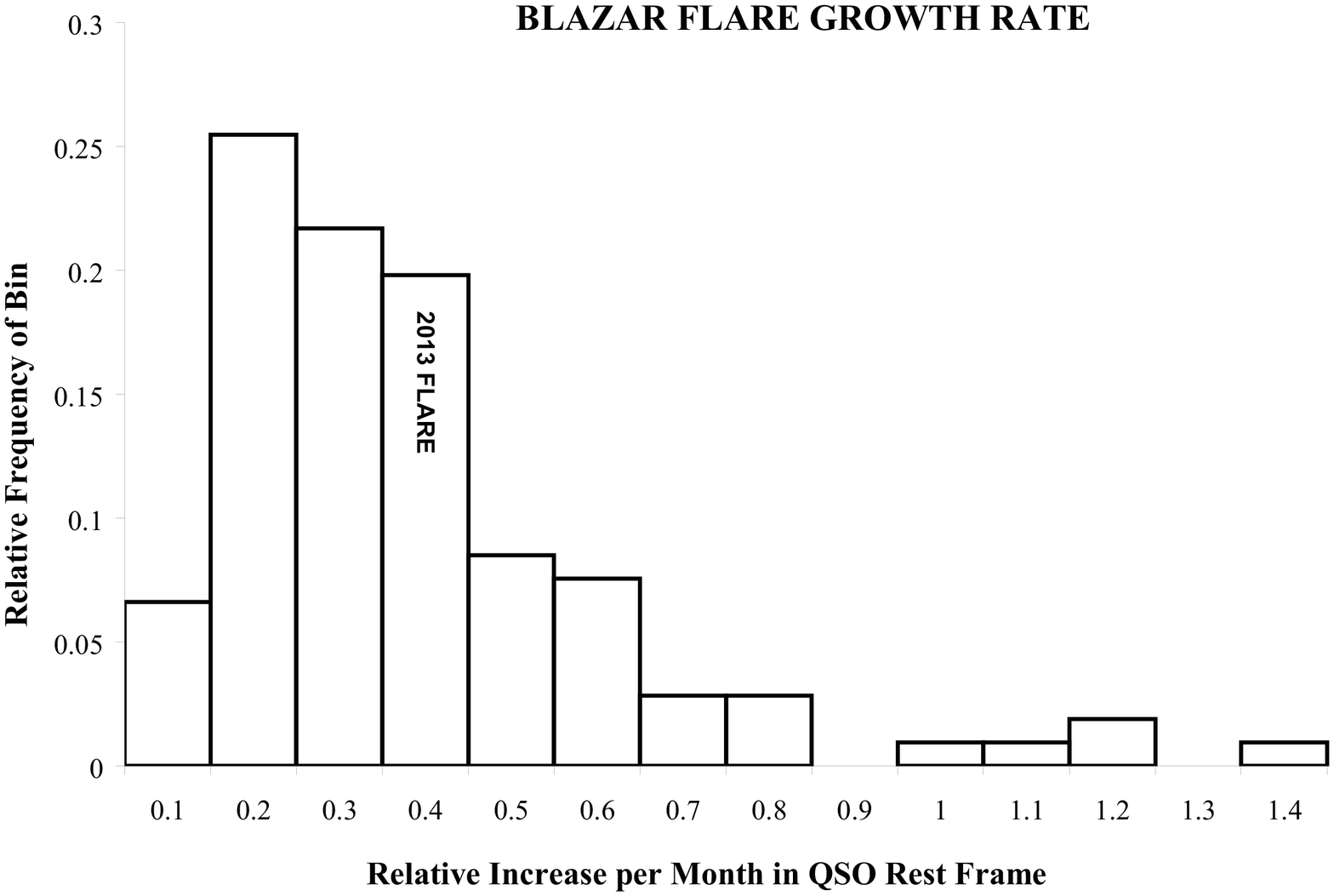}

\caption{Comparison of the 2013 flare in the Mrk\,231 to the
distribution of flare parameters in the \citet{ter04} blazar sample.
The top frame shows the distribution of the normalized flare
amplitude, $P/Q$. The bottom frame is the time rate of increase in
normalized units, $\dot{S}=(P/Q)$/(rise time), from the quiescent
baseline to the flare peak (units are $\mathrm{months}^{-1}$). Two
flares from \citet{ter04} are very abrupt and are off the bottom
histogram, to the right.}
\end{center}
\end{figure}
\subsection{Flare Definition} We define a flare as \emph{an abrupt
change in a light curve that results in a dramatic increase above
the quiescent background that precedes the flare.} There are two
quantitative components to describe this behavior: the abrupt change
that is expressed in terms of the time rate of increase of the light
curve, $\dot{S}$, and the large local maximum of the light curve,
$P$. These parametric descriptions are best expressed in units that
are normalized to the quiescent flux density level, $Q$, that
precedes the flare, i.e., the normalized amplitude $P/Q$ and
$\dot{S} = (P/Q)$/(flare rise time). The reason for the
normalization is the following. Consider a source with a quiescent
flux density of $ Q \sim 100$ mJy. A $\sim 100$ mJy increase in flux
density is a very significant relative change, $P/Q \approx 2$. By
contrast if the quiescent flux density is $ Q \sim 5 $ Jy, $\sim
100$ mJy increase in flux density is imperceptible, $P/Q \approx
1.02 $. A second critical aspect of defining a peak in the flux
density is that the local maximum be statistically significant above
the noise level of the quiescent background. This is important since
high frequency survey observations can have significant uncertainty
if the flux density is modest \citep{ter04}. The standard that we
adopt for defining a local maximum corresponding to a flare peak in
the light curve is that the putative peak is $P-Q >
3\sigma_{\mathrm{total}}$ above the quiescent baseline level, where
$\sigma_{\mathrm{total}}$ combines the uncertainty in the baseline
and peak measurements in quadrature, $\sigma_{\mathrm{total}}^{2} =
\sigma_{\mathrm{baseline}}^{2} +\sigma_{\mathrm{peak}}^{2}$.
\subsection{Defining the Quiescent Baseline}
A key element for defining the flare is the determination of
quiescent flux density that precedes the putative flare. This can be
difficult in general due to source flickering that is on the same
order of magnitude as the statistical uncertainty in each
measurement and occasionally a second flare occurs during the rise
of a previously initiated flare. Generally, in the latter
circumstance, these flares are not considered in our statistical
analysis since there is so much uncertainty created by this
circumstance. The only exception is if the second flare is clearly
much weaker than the first flare (i.e., it is essentially large
flicker noise). Our working definition of the baseline is a minimum
of 4 consecutive observations which agree within the 1$\sigma$
uncertainty and do not show a trend of increasing in time. This data
is linearly fit to determine the local (in time) baseline quiescent
level.
\subsection{The Radio Flares of Mrk\,231 in the Context of Blazars}
We compare the flare in Mrk\,231 to those of the flares discovered
by long term monitoring of $\sim $ 200 flat spectrum radio sources
at 22 GHz \citep{ter04}. We compare the flares in the 22 GHz light
curves to the flares in Mrk\, 231. Due to the condition defining a
statically significant flare, $P-Q > 3\sigma_{\mathrm{total}}$, all
flares satisfy $P/Q > 1.2$. Thus, we effectively segregate
traditional blazars from other less violently variable flat spectrum
AGN (non-blazars are rare in this high frequency selected sample).
The other implication of the $P-Q > 3\sigma_{\mathrm{total}}$
condition is that for noisy light curves, many flares cannot be
discerned cleanly from the noise using this standard and are
excluded from our analysis. A second complicating feature are gaps
in the temporal coverage that do not allow one of the following
three quantities to be determined, $Q$, $P$ or the start of the
flare. Other flares are excluded from our analysis due to a poor
determination of $Q$ that arises not only from gaps in sampling, but
noise superimposed on low flux density levels or uncertainty due to
two or more compound flares. In the end, we found 106 suitable
flares in \citet{ter04}. The distribution of $\dot{S}$ and $P/Q$ is
plotted in Figure 3. In order to compare $\dot{S}$ of objects at
different redshifts, the data were converted into the elapsed time
as measured in the QSO rest frame. The comparison of $\dot{S}$ and
$P/Q$ from the blazar sample and the Mrk\,231 flare shows that the
2013 flare is in-family (rise time and relative magnitude) with
blazar flares. If we consider the background contributions from
Equation~1, $S_{\nu=22\mathrm{GHz}}(\mathrm{core})$ tripled in 2013,
from $\approx 65$~mJy to $\approx 195 $~mJy in $\sim$ 150 days.
Based on two observations with VLBA, we can get crude estimates of
the parameters for the 2006 flare in Figure 1, $P/Q \sim 1.42 $,
$\dot{S} \sim 0.48$/month, also in-family with the blazar flares in
Figure 3.

\section{Conclusion}
In this letter, we demonstrated that strong,blazar-like, flares were
detected in the radio quiet quasar, Mrk\,231. Furthermore, this is
occurring in a BALQSO in which the radio jets are typically
suppressed. Ostensibly, Mrk\,231 is on the verge of becoming radio
loud, but the relativistic energy stream from the central engine is
being stifled by the BAL wind.
\par Another example of a strong radio flare
in a radio quiet quasar is III Zw 2. The magnitude of the flare in
1999 in this object is much more extreme. The 22 GHz flux increases
by a factor of $\sim$15 in 1.8 years \citep{ter04}. The amplitude is
off the chart in the top frame of Figure 3 and $\dot{S}=0.69$/month
is also larger than the 2013 flare for Mrk\,231. Such extreme
behavior occurs preferentially in radio quiet quasars since $Q$ is
generally small and the occasional large flare will be amplified by
this normalization. For example, the 1976 flare at 22.2 GHz in
Mrk\,231 discussed in \citet{rey09}, is very abrupt. It increased by
$134 \pm 37.5 $ mJy in 0.033 months, which leads to a relatively
large $P/Q= 2.33 \pm 0.65$ and an extremely large value of
$\dot{S}=70 \pm 20 $/month, which is off the chart in the bottom
frame of Figure 3. In spite of such dramatic behavior in these radio
quiet quasars there is one benign property of these flares that
indicates the radio quietness. The increase in the intrinsic radio
flux density of the flare (in the QSO rest frame) is very low
compared to the much higher redshift blazars in \citet{ter04}. Thus,
the variability brightness temperatures for the 2013 Mrk\, 231 and
the 1999 III Zw 2 flares are modest, $3.3 \times 10^{10}\,$ K and
$5.4 \times 10^{10}\,$ K, respectively, per the methods of
\citet{gho07}, $\sim 3$ orders of magnitude less than strong blazar
flares.
\par In conclusion, blazar-like flares are not rare in Mrk\,231, we detect one in various
stages of evolution half the time that we observe the source.
Further comparison to blazar flares can be made if superluminal
ejections are detected with VLBI observations that are triggered by
a flare detected with AMI monitoring. We currently have approved
VLBA observations to pursue this.

\section{Acknowledgments}
The National Radio Astronomy Observatory is a facility of the National Science
Foundation operated under cooperative agreement by Associated Universities,
Inc. This work made use of the Swinburne University of Technology software
correlator, developed as part of the Australian Major National Research
Facilities Programme and operated under licence. This research has made use of
NASA's Astrophysics Data System Bibliographic Services.


\begin{thebibliography}{}
\bibitem[Abraham et al.\ (1996)]{abr96} Abraham, Z., Carrara, E., Zensus, A.
      and Unwin, S. 1996, A\&A, 115, 543
\bibitem[Becker et al.\ (2000)]{bec00} Becker., R. et al.\ 2000, ApJ, 538, 72
\bibitem[Becker et al.\ (2001)]{bec01} Becker., R. et al.\ 2000, ApJS, 135, 227
\bibitem[Blundell et al(2003)]{blu03} Blundell, K., Beasley, A., Bicknell, G.
      2003, ApJL, 591, 103
\bibitem[Briggs et al.\ (1984)]{bri84} Briggs, F. H., Turnsheck, D. A., Wolfe,
      M. 1984, ApJ, 287, 549
\bibitem[Bruni et al.\ (2013)]{bru13}Bruni, G. et al.\ 2013 A\&A in press
      http://xxx.lanl.gov/abs/1304.3021
\bibitem[Brunthaler et al(2000)]{bru00}Brunthaler, A. et al.\ 2000, A\&A,
      357, L45
\bibitem[Edelson (1987)]{ede87}Edelson, R., 1987, ApJ, 313, 651
\bibitem[Ghosh and Punsly (2007)]{gho07}Ghosh, K. and Punsly, B. 2007, ApJL,
      661, 139
\bibitem[Gibson et al.\ (2009)]{gib09}Gibson, R. et al.\ 2009, ApJ, 692, 758
\bibitem[Hayashi et al.\ (2013)]{hay13}Hayashi, T., Doi, A., Nagai, H. 2013 to
      appear in ApJ http://xxx.lanl.gov/abs/1305.3371
\bibitem[Hewett and Foltz(2003)]{hew03}Hewett, P. and Foltz, C., 2003, AJ, 125,
      1784
\bibitem[Krichbaum et al.\ (1990)]{kir90}Krichbaum, T. 1990, A\&A, 237, 3
\bibitem[Kettenis et al.(2006)]{ket06} Kettenis,~M. et al.\ 2006, Astron. Data
      Anal. Software Syst. XV, 351, 497
\bibitem[Marscher et al.\ (1979)]{mar79} Marscher, A. et al.\ 1979, ApJ, 233,
      498
\bibitem[McCutcheon and Gregory (1978)]{mcc78}McCutcheon, W., Gregory, P. 1978,
      AJ, 83, 566
\bibitem[Mutel et al.\ (1990)]{mut90} Mutel, R. et al., Phillips, R., Su, B.,
      Bucciferro, R. 1990, ApJ, 352, 81
\bibitem[Punsly(2006)]{pun06}Punsly, B.\ 2006, ApJ, 647, 886
\bibitem[Reynolds et al.\ (2009)]{rey09}Reynolds, C., Punsly, B. Kharb, P.,
      O'Dea, C. and Wrobel, J. 2009, ApJ, 706, 851
\bibitem[Savolainen et al.\ (2002)]{sav02}Savolainen, T., Wiik, K., Valtaoja, E., Jorstad, S., Marscher, A. 2002, A\&A, 394,
      851
\bibitem[Shepherd et al.\ (1995)]{shepherd95}Shepherd, M., Pearson, T., Taylor,
      G. 1994, BAAS, 27, 903
\bibitem[Tateyama et al.\ (1999)]{tat99}Tateyama, C. et al.\ 1999, ApJ, 520,
      627
\bibitem[Ter\"asranta et al.\ (2004)]{ter04}Ter\"asranta, H. et al.\ 2004,
      A\&A, 427, 769
\bibitem[Ulvestad et al(1999a)]{ulv99}Ulvestad, J., Wrobel, J. and Carilli, C.
      1999, ApJ, 516, 134
\bibitem[Ulvestad et al(1999b)]{ulv00}Ulvestad, J. et al.\ 1999, ApJL, 517, L81
\bibitem[Weymann et al.\ (1991)]{wey91} Weymann, R.J., Morris, S.L., Foltz,
      C.B., Hewett, P.C. 1991, ApJ, 373, 23
\bibitem[Weymann(1997)]{wey97}Weymann, R. 1997 in ASP Conf. Ser. 128, Mass
      Ejection from Active Nuclei ed, N.Arav, I. Shlosman and R.J. Weymann (San
      Francisco: ASP), 3
\bibitem[Zhang et al.\ (2010)]{zha10}Zhang, S. et al.\ 2010, ApJ, 714, 367
\bibitem[Zwart et al.\ (2008)]{zwa08}Zwart, J. et al.\ 2008, MNRAS, 391, 1545
\end{thebibliography}
\end{document}